# Strongly Coupled Bacteriochlorophyll Dyad Studied Using Two-dimensional Phase-modulated Fluorescence-detected Electronic Spectroscopy

**VIVEK TIWARI,[1] YASSEL ACOSTA MATUTES,[2] ZHANQIAN YU,[3] MARCIN PTASZEK,[3] DAVID F. BOCIAN,[4] DEWEY HOLTEN,[5] CHRISTINE KIRMAIER,[5] ARKAPRABHA KONAR,[1] AND JENNIFER P. OGILVIE[1,\*]**

[1]*Department of Physics, University of Michigan, Ann Arbor, Michigan, 48109, USA*
[2]*Applied Physics Program, University of Michigan, Ann Arbor, Michigan, 48109, USA*
[3]*Department of Chemistry and Biochemistry, University of Maryland, Baltimore County, Baltimore, Maryland 21250, USA*
[4]*Department of Chemistry, University of California, Riverside, CA 92521, USA*
[5]*Department of Chemistry, Washington University, St. Louis, MO 63130, USA*
*\*jogilvie@umich.edu*

**Abstract:** Fluorescence-detected two-dimensional electronic spectroscopy (F-2DES) projects the third-order non-linear polarization in a system as an excited electronic state population which is incoherently detected as fluorescence. Multiple variants of F-2DES have been developed. However, none have demonstrated analysis of kinetics and coherences routine in photon-echo 2DES. Here, we report phase-modulated F-2DES measurements on a strongly coupled symmetric bacteriochlorin dyad, a relevant 'toy' model for photosynthetic energy and charge transfer. Coherence map analysis shows that the strongest frequency observed in the dyad is well-separated from the excited state electronic energy gap, and is consistent with a vibrational frequency readily observed in bacteriochlorin monomers. Kinetic rate maps show a picosecond relaxation timescale between the excited states of the dyad. To our knowledge this is the first demonstration of coherence and kinetic analysis using any variant of F-2DES.

## References and links

## 1. Introduction

Observations of coherent wavepackets in molecular aggregates such as light harvesting proteins[1, 2] and organic thin films[3] have sparked a growing experimental interest in exploring the role of coherent wavepackets in promoting energy and charge transfer[3-7]. These observations have also led to multiple theoretical models[8-10] which attempt to explain the origins of coherent wavepackets in terms of vibrational-electronic mixing.

There has also been a continued effort[4, 11-14] to understand how coherent wavepackets resulting from purely vibrational Franck-Condon (FC) displacements, purely electronic Coulomb coupling or vibrational-electronic mixing manifest in a two-dimensional electronic spectroscopy[15] (2DES) experiment. To that end, several recent 2DES experiments on considerably less complex 'toy' models of electronically coupled chromophores have been reported. However, artificial dimers, such as in refs.[16-19], employed in order to emulate a photosynthetic dimer have electronic couplings and Huang-Rhys factors an order of magnitude larger[20] than what is known in photosynthetic proteins, potentially exploring altered mechanisms of energy and charge transfer than those proposed to occur in photosynthetic proteins[10]. In order to explore the same physics as is involved in photosynthetic energy and charge transfer in a toy model, weakly coupled FC active vibrations such as those present in chlorophylls and bacteriochlorophylls[21-24] are essential.

A 2DES spectrum resolves the response of a system to pump and probe pulses on a correlated contour map with excitation and detection axes. By resolving the dynamics along waiting time $T$ into 2DES coherence and rate maps[11, 12, 25], information about the physical origin of the coherence (vibrational versus mixed vibrational-electronic) and the identification of the participating states can be routinely deduced in a variety of systems[4, 12-14]. 2DES experiments measure the four-wave-mixing response from an ensemble of molecules and are sensitive to inhomogeneities in the ensemble, which affect the waiting time dynamics[10, 26, 27]. Conclusions regarding the possible survival of electronic wavepackets in individual members of the ensemble based on experimentally observed coherence dephasing timescales, are amenable to ensemble dephasing effects. The need for probing at sub-ensemble spatial resolution with high sensitivity has motivated the development of fluorescence based 2DES experiments, first demonstrated by Warren and co-workers[28]. All collinear geometry and fluorescence-based detection holds promise for probing sample sizes smaller than $\lambda^3$, obviating the requirement of generating a phase-matched signal from a macroscopic grating of oscillating dipoles. While multiple approaches[28-31] to fluorescence-detected 2DES spectroscopy (F-2DES) have been developed, none have demonstrated the feasibility of the approach for analysis[11, 12, 25] of the coherent waiting time dynamics routinely done in conventional 2DES experiments.

Here, we report F-2DES measurements of weakly underdamped coherent wavepackets in a strongly coupled symmetric bacteriochlorin dyad[32]. We adopt the phase-modulated approach to F-2DES developed by Marcus and co-workers[29]. The electronic couplings and FC displacements[24, 33-36] deduced from the absorption spectrum of the dyad, closely mimic those of a typical photosynthetic protein. Here we use this model system to demonstrate the feasibility of F-2DES spectroscopy for coherence and kinetic analysis[11, 12, 25] of coherent wavepackets performed routinely in 2DES experiments.

## 2. Theory

A 2DES experiment is comprised of three light-matter interactions with precisely controlled time intervals between them. The three interactions create a third-order macroscopic polarization $\hat{P}^{(3)}(\vec{k}_s, t, t_a, t_b, t_c)$ in the system, where $t_a, t_b, t_c$ represent the experimentally controllable time points at which the non-collinear pulses $a$, $b$, $c$ are incident on the sample, and $t$ denotes the time after the arrival of pulse $c$. The oscillating polarization radiates an electric field along a background free phase-matched direction determined by the wave-vectors of the three incident pulses, that is, $\vec{k}_s = -\vec{k}_a + \vec{k}_b + \vec{k}_c$. In an F-2DES experiment, the pulses are typically collinear, and an additional pulse $d$ projects the oscillating polarization $\hat{P}^{(3)}(\vec{k}_s, t, t_a, t_b, t_c)$ as a population on an excited electronic state. The fluorescence from such a population contains the phase information encoded by the four light-matter interactions. Additionally, in F-2DES, the use of a fully collinear geometry enables integration of the setup with a microscope for spatially-resolved measurements[37, 38]. The spatial resolution of such measurements could be enhanced from ~diffraction-limited resolution to the nanoscale with the use of nanoantennas[39] for excitation localization. With fully collinear pulses, signal detection in a background free direction is no longer possible. This drawback has been either overcome by phase-cycling, first demonstrated by Warren and co-workers[28], or a phase-modulation approach to F-2DES developed by Marcus and co-workers[29]. The current study utilizes the phase-modulation approach which will be described in detail in the following Section.

Wave-mixing diagrams equivalent to those which describe the evolution of density matrix elements in a 2DES experiment, are also valid for an F-2DES experiment. However, as noted by Marcus and co-workers[40], the excited state absorption (ESA) signals in F-2DES can have a negative as well as a positive sign relative to the ground state bleach (GSB) and excited state emission (ESE) signals. In contrast, in a 2DES experiment, the ESA signal is always negatively-signed relative to the GSB and ESE signal contributions. The wave-mixing pathways corresponding to the GSB, ESE and ESA signals from a three electronic level system are shown in Fig. 1A. A positive sign for ESA results in signal additions instead of cancellations known in 2DES. This can result in strikingly different 2D peak shapes in F-2DES spectrum of certain systems[40-42].

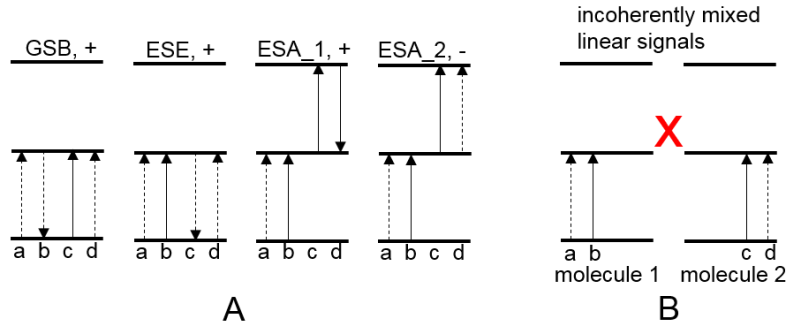

Fig. 1. (A) Wave-mixing diagrams which contribute in F-2DES experiment on a three electronic level system. The dashed and solid vertical lines represent successive field-matter interactions with laser pulses a-d, from the bra and ket side of the density matrix element, respectively. All pathways correspond to rephasing 2D pathways. + and − denote the relative signs of the signal contributions. Unlike 2DES, ESA signal has a positive sign (ESA_1). The ESA_2 signal fluoresces from the doubly excited state, where exciton-exciton annihilation and non-radiative relaxation within a denser manifold electronic states reduces the fluorescence quantum yield causing weak to no signal contributions[40] from ESA_2 type paths. (B) A diagrammatic depiction of incoherent mixing (represented by red 'X') of linear fluorescence signals[43] in F-2DES, between two uncorrelated molecules represented by different ground state for each molecule. A possible scenario where linear fluorescence signal from pump and probe pulses can mix incoherently involves energy transfer from molecule 1 to 2, the probability of which is determined by the Förster critical distance[44].

Because F-2DES relies on incoherent fluorescence detection, mixing of *linear* population signals during the detector integration time $t$ is possible for certain systems. For example, Silva and co-workers[43] have shown that, in semiconductors with high excited carrier densities, mixing of linear population signals from the first and second pair of light-matter interactions, through pulse pairs *a, b* and *c, d*, respectively, can occur due to carrier annihilation processes such as Auger recombination. Such incoherently mixed signals can have the same phase information as the coherent non-linear signals, and hence could be hard to distinguish through fluorescence-detection. Fig. 1B shows that pathways corresponding to incoherently mixed contributions are equivalent to the coherent ESE pathways, which create a population after the first two interactions. Note that in order for the signal to be considered incoherent, the first pair and second pair of four light-matter interactions should happen on two uncorrelated systems. This is represented by two separate ground states Fig. 1B, that is, two different dyad molecules. Such a scenario may be possible if energy transfer between two molecules is followed by exciton-exciton annihilation. The Forster critical concentration[44] for energy transfer from one molecule to another is of the order of mM, for which the critical transfer distance between the molecules is ~50 Å or less. The concentrations used in the present study are ~8 μM, nearly $10^3$ times below the critical concentrations for energy transfer between the molecules, making such incoherent contributions unlikely in our experiment.

## 3. Experiment

### 3.1 Experimental Setup

The experimental layout, shown in Fig. 2, is based on the original design by Marcus and co-workers[29]. The pulse train from a broadband 83 MHz Ti:Sapphire oscillator (Venteon One) is routed into a SLM-based pulse shaper (MIIPS 640P, Biophotonic Solutions) for dispersion pre-compensation. Laser pulses from the pulse shaper (19 fs FWHM pulse duration) are split by a 50:50 beamsplitter (BS, Newport, 10B20BS.2), and each portion is routed into two Mach-Zehnder (MZ) interferometers. Within each MZ, the pulse is further split using 50:50 BS, and each arm is tagged with a unique carrier-envelope phase which is cycled at radio frequencies using an acousto-optic modulator (AOM, Isomet M1142-SF80L). The phase modulation frequencies for the four arms are denoted as $\Omega_{i=1-4}$. When the split pulses are recombined at BS3 and BS5, within MZ1 and MZ2 respectively, the pulse amplitude modulates at the difference frequencies of the AOMs, that is, $\Omega_{12}$ and $\Omega_{34}$ for MZ1 and MZ2, respectively. One of the two output ports from BS3 and BS5 is routed to a monochromator (Optometrics, MC1-04) to generate the reference frequencies, $\omega_1^R = 12422$ cm$^{-1}$ (805 nm), and $\omega_2^R = 12422$ cm$^{-1}$ (805 nm), for MZ1 and MZ2 respectively. The references generated from MZs 1 and 2 are used for phase-sensitive lock-in detection[29], while the second output ports from MZ1 and MZ2 are recombined at BS6.

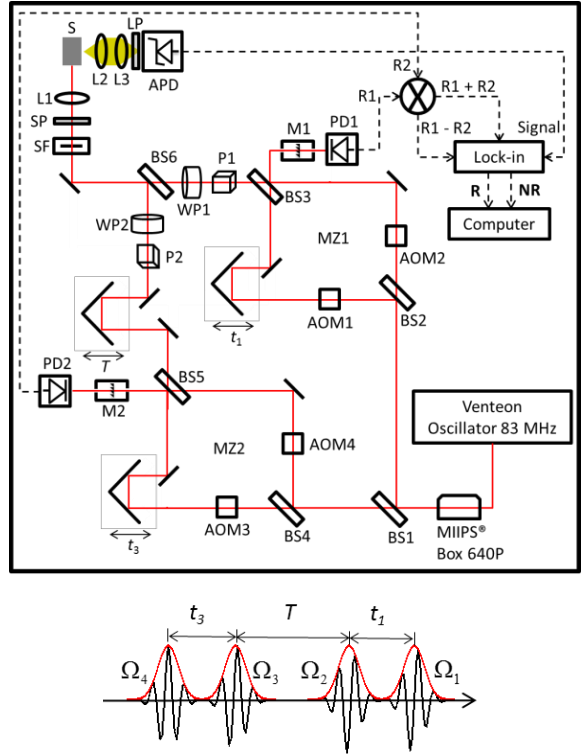

Fig. 2. Experimental setup of the F-2DES spectrometer. Beamsplitter (BS), Acousto-Optic Modulator (AOM), Mach-Zehnder interferometer (MZ), Monochromator (M), Polarizer (P), Waveplate (WP), Photodiode (PD), Spatial Filter

(SF), Shortpass Optical Filter (SP), Lens (L), Sample (S), Longpass Optical Filter (LP), Avalanche Photodiode (APD), Electronic Reference Signal (R1,R2), Rephasing (**R**) and Nonrephasing (**NR**) 2D signals. The experimentally controllable carrier envelope time delays $t_1, T, t_3$, between pulses a-b, b-c and c-d in Fig. 1, respectively, are shown in the lower panel.

The time delays between all the collinear pulses resulting after BS6, are controlled by translational stages (Newport M-VP25XL for $t_1$ and $t_3$ time delays, and ILS150 for $T$ time delay), where $t_1$ is the time delay between the pulses *a* and *b*, $t_3$ is the time delay between pulses *c* and *d*, and $T$ is the delay between the pulses *b* and *c*. Typically, $t_1$ and $t_3$ delays are scanned from 0 to 90 fs in steps of 5 fs, and the $T$ delay is scanned from 0 to 2 ps in steps of 10 fs. The signal level at $t_{1,3} = 90$ fs is below 5% of the signal at $t_{1,3} = 0$ fs at room temperature.

The four-wave-mixing signal generated by the sample oscillates at the difference frequencies $(\Omega_3 - \Omega_4) \pm (\Omega_1 - \Omega_2)$, where the (positive) negative sign corresponds to (non-rephasing) rephasing 2D signals. The oscillating signal is demodulated using a lock-in amplifier (Zurich Instruments, HF2LI) with phase-sensitive detection. The reference signals $\omega_1^R$ and $\omega_2^R$, from monochromators 1 and 2 respectively, are mixed in a 24 bit digital signal processor (Analog Devices, ADAU1761) to generate the reference signals modulating at frequencies $\Omega_{34} \pm \Omega_{12}$. These reference signals are connected to the lock-in amplifier reference channels corresponding to rephasing and non-rephasing signals. Physical undersampling of the oscillating signal is achieved by signal detection relative to the reference frequencies, $\omega_1^R$ and $\omega_2^R$ from MZ1 and MZ2 respectively. That is, if the signal oscillates at frequency $\omega_{eg}$ corresponding to the electronic energy gap of the molecule during $t_1$ and $t_3$, then the physically undersampled signal[29] oscillates at frequency $\omega_{eg} - \omega_{1,2}^R$. For the experiments, the AOM frequencies are set at $\Omega_1$ = 80.111 MHz, $\Omega_2$ = 80.101 MHz, $\Omega_3$ = 80.029 MHz and $\Omega_4$ = 80.0 MHz through a common clock (Novatech DDS Model 409B), such that the resulting signals oscillate at 19 kHz (rephasing) and 39 kHz (non-rephasing).

The train of four collinear phase-modulated pulses is focused in a 1 mm pathlength sample flowcell (Starna 584.4-Q-1) using a 5 cm focusing lens (L1) with a FWHM focal spot size of ~16 μm. The resulting fluorescence is collected at 90° using an 815 nm longpass filter (Chroma) mounted on an APD (Thorlabs APD 110A). The details of the dyad synthesis are described in ref. [32]. The sample (10 ml typical volume, OD ~ 0.16 in 1mm cuvette) was prepared and stored in a nitrogen glovebox. The solvent (99.8% anhydrous Toluene, Sigma) for preparing the sample was permanently stored in the glovebox. The sample was circulated through the cuvette using a peristatic pump (Masterflex Model 07516-00) at average flow rate of ~260 ml/min. The spectra are collected with 3 pJ pulse energies, such that the pump excitation probability in front of the sample at the center of the pulse is 0.4%. Low excitation probability is chosen so as to minimize multiple excitations on the molecule due to sample flow rates not being fast enough to counter the 83 MHz repetition rate.

### 3.2 Comparison with phase-cycling approaches

Using a grating based acousto-optic pulse shaper, Warren and co-workers have demonstrated F-2DES on rubidium atoms. Their 16-step phase cycled approach is scalable to higher laser repetition rates. More recently, F-2DES has also been demonstrated[30, 31, 37] using a 27 step phase-cycling approach for filtering out the non-linear fluorescence signal. Each of the 27 phase cycles are performed by either an acousto-optic pulse-shaper (Dazzler, Fastlite)[31], or a spatial light modulator based pulse shaper[30, 37]. At each step, a different acoustic waveform imparts a set phase to each of the four identical pulses generated by the pulse-shaper, and the filtered signal is obtained by adding the signals resulting from each of the 27 phase cycles. Typically, the refresh time for the waveforms are of the order of a few milliseconds, thus limiting the data collection time, and making the phase-cycling process susceptible to laser noise and any possible sample photobleaching. Pulse amplitude shaping approaches[31] may also constrained by the pulse-shaper to repetition rates of a few kHz, whereas typically repetition rates of the order of MHz are optimal for fluorescence detection of single molecules. Additionally, independent spectral[45] and polarization[46] controls over all the four pulses, typically done for enhancing certain features in 2DES experiments, are not easily implementable. In contrast, the phase-modulation approach to F-2DES has been demonstrated[40] down to 250 kHz repetition rates with easy accessibility to MHz repetition rates. Each successive pulse from the oscillator pulse train is tagged with a unique carrier envelope phase which cycles at the AOM frequency $\Omega_{i=1-4}$. Thus, the phase-cycling is performed in real time as the pulse train from the oscillator passes through the AOMs. Using independent AOMs in each arm of the interferometer has an added advantage of enabling independent control over the spectrum and polarization of each pulse. Of all the F-2DES approaches mentioned above, only Draeger et al.[31] have reported coherent wavepackets in a laser dye, along waiting time $T$ at room temperature.

For $t_1$ and $t_3$ step sizes of 5 fs, the total number of $t_1$ and $t_3$ steps per $T$ are 19x19 = 361 steps, for each absorptive 2D spectrum presented in the paper. For 361 time steps per $T$, the total data collection time per 2D spectrum, including the stage waiting time, lock-in electronics and data acquisition, is ~48 seconds. In the phase-cycling approach in ref.[31], with 225 time steps (0 to 84 fs in 6 fs step) per 2D spectrum and 10 averages to reach the signal-to-noise ratio analyzed in the paper, the data collection time per 2D spectrum is 60 seconds. However, in the phase-modulation approach, with reference wavelength at 805 nm and the upper electronic state in the dyad at 775 nm, the largest undersampled frequency sampled by $t_1$ and $t_3$ scans corresponds to ~481 cm$^{-1}$. Therefore, sampling in 5 fs steps implies that a maximum frequency of 3335 cm$^{-1}$ could be sampled without aliasing artifacts. In other words, sampling in larger $t_1$ and $t_3$ step sizes is possible without introducing any artifacts. For example, Fig.A1 in the appendix shows that, scanning from 0 to 90 fs in steps of 10 fs reduces the total number of time steps to 100 steps and the data collection time per 2D spectrum to only ~12 seconds, without affecting the data quality.

## 4.  Results and Discussion

Fig. 3A shows the chemical structure of the symmetric bacteriochlorin dyad, BC$_2\beta$E[32]. Fig. 3B shows the linear absorption spectrum of the bacteriochlorin dyad along with the laser spectra overlaid in red and blue. The linear absorption spectrum of the dyad shows a prominent electronic splitting of ~580 cm$^{-1}$. Bocian and co-workers[32] have explained the two electronic transitions in the dyad in terms of electronic states resulting from the linear combination of atomic orbitals (LCAO) of the frontier orbitals in the bacteriochlorin monomers. Strong electronic interactions on the molecular orbitals of the dyad result in an electronic splitting of 580 cm$^{-1}$ in the $Q_y$ band. The Huang-Rhys factors for the FC active vibrations in bacteriochlorin monomers are expected to be of the order of 0.05[22, 23]. Thus, the electronic coupling as well as the Huang-Rhys factors are within the parameter range known for chromophores in various photosynthetic proteins[47]. Resonance Raman experiments and normal mode analyses on bacteriochlorin derived molecules[33-36] suggest that a prominent 580 cm$^{-1}$ FC active mode arises due to pyrrole and pyrroline ring deformations. Assuming that a similar mode is also present on the bacteriochlorin skeleton, non-adiabatically enhanced vibrational wavepackets due to a vibrational-electronic resonance might be expected in this system[10]. However, since the present dyad is not an excitonic dimer simply involving excited-state transition dipole-dipole coupling and has mixed molecular orbitals, split absorption features and redox potentials[32], deviations in excited-state properties from a non-adiabatic excitonic dimer[10] should not be surprising. Fig. 3C shows the absorptive F-2DES spectra for $T = 0$ and 2 ps.

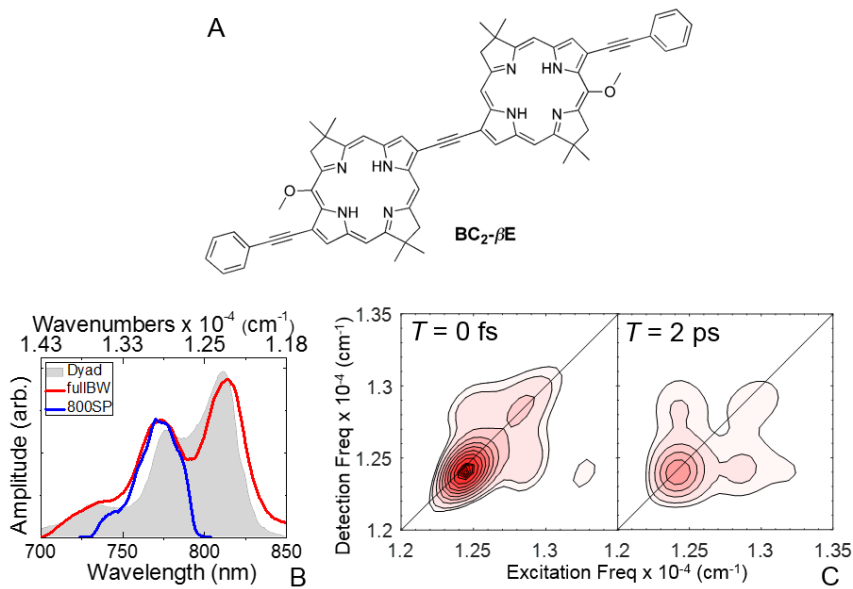

Fig. 3. (A) Chemical structure of the symmetric bacteriochlorin dyad $BC_2\beta E$. (B) Linear absorption spectrum of the dyad with the full laser spectrum ('fullBW'), and the laser spectrum with an 800 shortpass filter ('800SP'). (C) Absorptive F-2DES spectra at waiting time $T = 0$ and 2 ps, corresponding to the 'fullBW' laser spectrum. Contours are drawn at 5%,10-90%,95%,98% and 100% signal levels.

Based on the wave-mixing diagrams similar to those in Fig. 1A, refs.[41, 42] argue that presence of well-resolved cross-peaks (CPs) between the electronic states at $T = 0$ fs implies delocalized excited electronic states which share a common ground and doubly-excited state. Ignoring CP contributions with non-perfect FC overlap factors, only GSB and ESA wave-mixing paths contribute to non-oscillatory $T = 0$ fs CPs. Note that the dominant ESA signal contribution in F-2DES is positive[29], compared to negative in 2DES spectroscopy, which can give rise to qualitatively different looking 2D spectra[40, 41].

Figs. 4A,B show the 'pump-probe' scan obtained by scanning $T$ from -10 ps to 800 ps. Signal for $T < 0$ fs is expected in F-2DES because the conjugate pulse pairs (a,b) and (c,d) are interchangeable[48]. The results of the three-exponential fit to the 'pump-probe' data are shown in Table 1. In the case of '800SP' excitation, when only the upper electronic state is excited, the fastest decay component of 1.2 ps is ~24x larger in amplitude. Fig. 4C shows the 1.1 ps kinetic rate map obtained by fitting the 2D dataset obtained by scanning $T$ from 0 fs to 2 ps, to a global bi-exponential decay[49]. The picosecond decay rates obtained from 'pump-probe' data and the full 2D data are within the error bars from the fit.

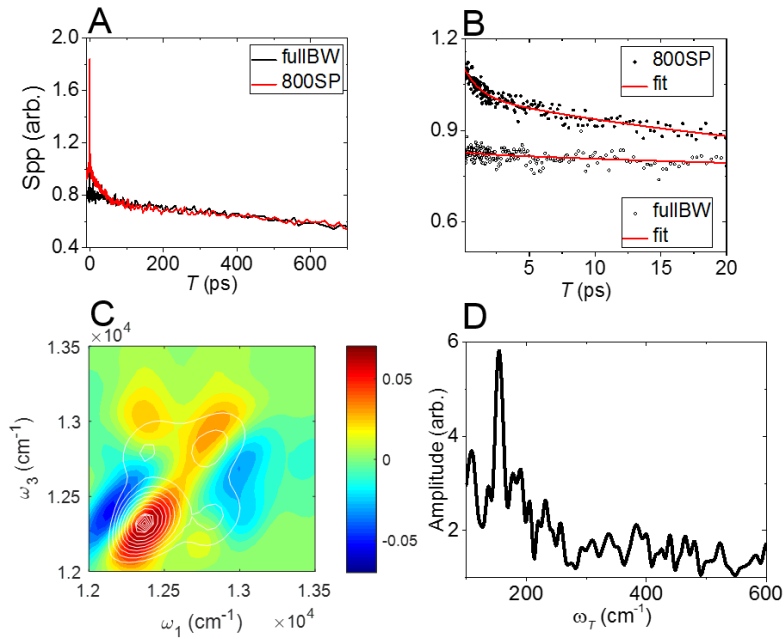

Fig. 4. (A) 'Pump probe' scan obtained by fixing $t_{1,3} = 0$ fs, and scanning $T$ from -10 ps to 800 ps. The minimum step size is 20 fs and the step size doubles after every 2 ps. The 'fullBW' case corresponds to the red laser spectrum in Fig. 3B, whereas '800SP' corresponds to the blue laser spectrum, where only the upper electronic state is pumped and probed. The scans are normalized to 800ps. (B) A global fit of both the scans of panel (A) to a three exponential decay. The results of the fit are shown in Table 1. When only the upper electronic state is excited, a ~24x larger amplitude is seen in the fastest (1.2 ps) decay. (C) 2D rate map for 1.10 (± 0.14) ps time constant derived by globally fitting the $T$ evolution of all pixels in the F-2DES spectra from $T = 10$ fs to 2 ps. The pixels were fit to a bi-exponential decay with an offset. Negative (blue) regions correspond to exponential growth, and positive (red) regions correspond to exponential decay with 1.10 ps time constant. The faster exponential decay from the fit was 21.9 (± 0.9) fs and corresponds to the pulse overlap region. The sum-of-squares residual norm from the global fit was 0.022. Absorptive $T = 10$ fs F-2DES spectrum is overlaid in the background for reference. (D) Frobenius spectrum obtained from the 2D dataset after subtracting the global exponential fits. The strongest frequency in the data corresponds to ~155 cm$^{-1}$, while other less prominent modes at 110 cm$^{-1}$, 180 cm$^{-1}$ and 230 cm$^{-1}$ are also present.

Table 2. Results of fitting the 'pump-probe' data in Fig. 4B.

| | A1 x 10$^2$ | T$_1$ (ps) | A2 x 10$^2$ | T$_2$ (ps) | A3 x 10$^2$ | T$_3$ (ns) |
|---|---|---|---|---|---|---|
| **fullBW** | 0.4 (±0.6) | 1.2 (±0.1) | 4.9 (±0.4) | 30.2 (±1.4) | 77.4 (±0.3) | 2.35 (±0.06) |
| **800SP** | 9.5 (±0.7) | | 25.7 (±0.5) | | 75.5 (±0.3) | |

The apparent absence of a large picosecond amplitude for 'fullBW' case in Figs. 4A,B is explained by the presence of growing and decaying regions seen in the 2D rate map of Fig. 4C, which will partially cancel once the 2D spectrum is integrated along both axes to obtain a pump-probe trace.

Fig. 4D shows the oscillatory content in the 2D data obtained by subtracting the exponential background, and calculating a Frobenius norm of the entire 2D spectrum. The strongest mode seen in the data is ~155 cm$^{-1}$. A FC active vibrational mode at this frequency is well established for bacteriochlorin-derived molecules, and arises on the bacteriochlorin macrocycle skeleton[33, 36, 50]. Oscillations of 2D peak at this frequency have been consistently reported[1, 2, 14, 51] across a variety of photosynthetic proteins with bacteriochlorin-derived chromophores. Surprisingly, no prominent mode is observed near the electronic splitting frequency of 580 cm$^{-1}$. This suggests two possibilities: 1. No detectable purely electronic coherence is present at room temperature in the measured ensemble of dyads, or 2. Possible absence of non-adiabatically enhanced vibrational wavepackets[10]. This will be a subject of future F-2DES and resonance Raman experiments on bacteriochlorin monomers.

The longest time constant of 2.35 ns obtained from the 'pump-probe' data in Fig. 4A,B agrees well with the previous determination[32] of 2.1 ns population lifetime of the dyad in toluene. Previous studies[32, 52] on symmetric bacteriochlorin dyads have reported a quenching of fluorescence in polar solvents caused by increased charge-resonance (CR) character of the lowest $Q_y$ electronic state of the dyad in a polar solvent. The increased CR character in the lowest electronic state allows fast internal conversion to the ground electronic state. Such a quenching has been reported to be on a timescale of tens of picoseconds. The 30 ps timescale seen in the 'pump-probe' data is close to the expected[52] timescale of internal conversion to the ground electronic state. However, internal conversion due to increased CR character has only been reported in polar solvents such as DMF, and was found to be absent in toluene. The presence of a ~30 ps relaxation component even in non-polar solvents such as toluene hints at some degree of CR character of the lowest $Q_y$ electronic state, even without a polar solvent[32]. In Fig. 4C, a picosecond decay on the upper DP region is accompanied by a simultaneous rise in the lower CP region, indicating that the fastest ~1 ps component is the relaxation from the upper to the lower $Q_y$ electronic state. However, for the pump excitation frequency corresponding to the lower electronic state, a picosecond decay is seen at the upper upper CP, as well as decay/growth patterns on the lower electronic state. The above observations hint at the presence of another relaxation process on the picosecond timescale – population transfer to another state energetically close to the lower electronic state leading to a concurrent growth of ESA of the transferred population. A concurrent growth of the ESA signal is well-known in certain photosynthetic proteins[53]. Solvent polarity dependence of 2D rate maps could provide further insight into this hypothesis.

Fig. 5A resolves the Frobenius spectrum of Fig. 4D on a 2D coherence map. Separation of coherent signal into positive and negative coherence frequency is routinely [11-13] done in 2DES experiments in order to determine the physical origin (ground versus excited state, and purely vibrational, electronic or vibrational-electronic) of the coherent signal. The absolute value rephasing coherence maps for the most prominent coherence frequency $\omega_T = \pm 155$ cm$^{-1}$ are shown on the same contour scale. Fig. 5A shows that the oscillation amplitude for this vibration is primarily localized on the lower electronic state, with only ~15-20% amplitude on the upper electronic state and the cross-peak. The transition strength to the upper electronic state is ~64% of the transition strength to the lower electronic state (Fig. 3B). Therefore, vibrational modulation amplitude on the upper excited electronic state is expected to be only ~40% compared to the lower electronic state. The amplitude is expected to be further lowered to ~20% compared to the lower electronic state because the laser amplitude on the upper electronic state is ~73% of the laser amplitude on the lower electronic state (Fig. 3B), thus accounting for very weak oscillation amplitude on the upper electronic state. As seen in Fig. A2 in the appendix, the observed vibration is weakly underdamped and survives on a picosecond timescale. Such underdamped vibrational wavepackets are known[54] to not directly contribute to vibrational-electronic mixing assisted energy or charge transfer, although their presence could hint at the aftermath of energy transfer in certain cases[10].

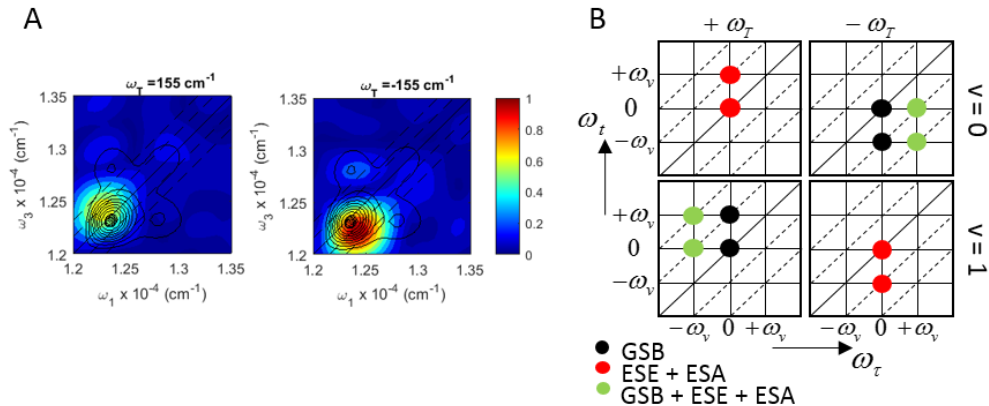

Fig. 5. (A) Absolute value rephasing coherence maps for $\omega_T = \pm 155\,\text{cm}^{-1}$. The dashed line are displaced from the diagonal by $\pm 155\,\text{cm}^{-1}$. Real absorptive 2D spectrum at $T = 50$ fs is overlaid in the background for reference. (B) Contributions to the rephasing F-2DES coherence map expected from a displaced harmonic oscillator model for an isolated pigment with one FC active vibrational mode with frequency $\omega_T$. The model predicts purely vibrational coherences at frequency $\omega_T$ which are separable into positive and negative coherence frequencies shown in the left and right columns, respectively. The upper panels correspond to contributions for which the corresponding Feynman pathways start from the zero-point level (v = 0) on the ground state. Lower panels show contributions for which the corresponding Feynman pathways start from v = 1 vibrational level on the ground electronic state. In F-2DES, ESA signal can have positive and negative contributions. Assuming that electronic state with more than one quantum of excitation have negligible quantum yields, the ESA signal considered in this analysis has a positive sign, similar to GSB and ESE signals.

Fig. 5B shows the expected locations of vibrational coherences for a rephasing F-2DES absolute value coherence map plotted for positive and negative coherence frequencies. For the 155 cm$^{-1}$ vibrational mode, assuming all other vibrations are in v = 0 level, the population in v = 1 vibrational level is approximately half compared to that in v = 0 at 300 K, and therefore its contribution to the coherence maps is *not* expected to be negligible. Experimentally, the coherence map contributions from such low-frequency vibrations are not expected to be well-resolved[55]. Hence, the theoretically separable contributions shown in Fig. 5B are experimentally merged together for low frequency modes at 300 K. The analysis in Fig. 5B shows that for *negative* coherence frequencies, coherences from both ground and excited state contribute from v = 0, and only weak excited state coherences contribute from a hot ground state (v = 1 pathways). Overall there are a larger number of strong contributions for negative coherence frequency, expectedly giving rise to a stronger contribution compared to positive coherence frequency.

## 5. Conclusion

We have demonstrated the first F-2DES analysis of kinetic rate maps and coherence maps on a strongly-coupled bacteriochlorin dyad at room temperature. We do not detect any signatures of purely electronic coherence. A vibrational-electronic resonance is energetically established for the dyad, but no signatures of enhanced vibrational wavepackets at the resonant vibrational frequency were found. This may indicate that appropriate symmetry requirements for the enhancement of vibrational wavepackets are not met. It may also reflect the deviation of the dyad from a purely excitonic dimer[32]. A picosecond transfer between the excited electronic states of the dyad and weakly underdamped low-frequency vibrational wavepackets were observed. Future F-2DES and resonance Raman experiments on bacteriochlorin monomers, and solvent dependence of the rate maps will shed more light on the reported findings.

**Funding**

The work at UM was supported by the AFOSR Biophysics program under grant FA9550-15-1-0210. Y. A. M. also acknowledges the Rackham Merit Fellowship program at the University of Michigan. Work at WU and UCR was supported by grants from the Division of Chemical Sciences, Geosciences, and Biosciences, Office of Basic Energy Sciences of the U.S. Department of Energy, DE-FG02-05ER15661. M.P. was supported by NSF (CHE-1301109).

**Acknowledgments**

The authors thank Julia Widom, Andrew Marcus, and Eric Martin, Travis Autry of the Cundiff group for their enthusiastic help in building the F-2DES setup.

**Appendix**

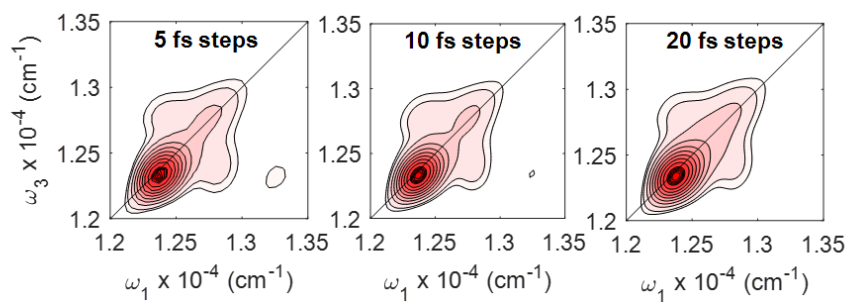

Fig. A1. Absorptive F-2DES spectra at waiting time $T = 0$ fs, for $t_{1,3}$ step sizes of 5 fs (361 points, ~48 secs collection time), 10 fs (100 points, ~12 secs collection time) and 20 fs (25 points, ~5 secs collection time). The original 2D spectrum was collected from 0 to 90 fs in steps of 5 fs (same as shown in Fig. 3B), while the 10 fs and 20 fs spectra are obtained by ignoring the extra time points. For the 2D spectrum with 20 fs time steps, the data was truncated at 80 fs where the signal levels are >10%. Hence, for that case, the peaks appear relatively broader than other cases. As mentioned in the Section 3.2, with the reference wavelength at 805 nm and the upper electronic state in the dyad at 775 nm, the largest undersampled frequency sampled by $t_{1,3}$ scans is ~481 cm$^{-1}$. The Nyquist limit is 3335 cm$^{-1}$ for 5 fs step size, and ~833 cm$^{-1}$ for 20 fs step size. Hence, physical undersampling in F-2DES allows rapid data collection without additional averaging or compromising on peak amplitudes, position or aliasing artifacts, as long as the step size is sufficiently away from the Nyquist limit. Contours are drawn at 5%, 10-90%, 95%, 98% and 100% signal levels.

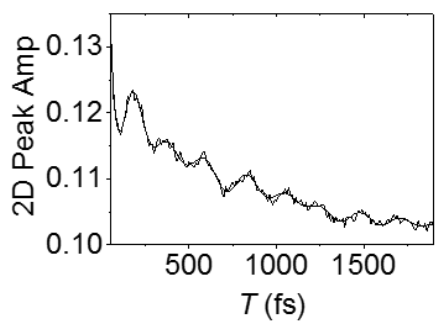

| Freqs (cm⁻¹) | Damping (fs) | +/- Error (fs) | Amp x 1E-4 |
|---|---|---|---|
| 189 | 116 | 42 | 86.3 |
| 152 | 1215 | 296 | 20.6 |
| 108 | 853 | 203 | 18.2 |
| 51 | 1900 | 1183 | 5.37 |

Fig. A2. (Top) Raw $T$ trace of the real rephasing 2D spectrum at the pixel corresponding to the maximum coherence amplitude for coherence frequency $\omega_T = 155 \text{ cm}^{-1}$. The raw trace is overlaid with a global damped sinusoidal fit from 9 pixels around the maxima. The fit function consists of an offset, two non-sinusoidal exponential rates corresponding to Fig. 4C, and four exponentially decaying sinusoids with their respective phases. The globally shared parameters are the exponential decay rates and the damping rates, while the offsets, amplitudes and phases are allowed to float for each pixel. Fits with higher than four sinusoids did not give a smaller sum of squares residual. The results of the fit are shown in the accompanying table, and show a room temperature damping timescale of a picosecond for the 155 cm⁻¹ vibrational wavepackets.